\documentstyle[12pt]{article}
\def\PRL #1 #2 #3{{\sl Phys. Rev. Lett.} {\bf#1} (#2) #3}
\def\NPB #1 #2 #3{{\sl Nucl. Phys.} {\bf B #1} (#2) #3}
\def\NPBFS #1 #2 #3 #4{{\sl Nucl. Phys.} {\bf B #2} [FS#1] (#3) #4}
\def\CMP #1 #2 #3{{\sl Commun. Math. Phys.} {\bf #1} (#2) #3}
\def\PRD #1 #2 #3{{\sl Phys. Rev.} {\bf D #1} (#2) #3}
\def\PLA #1 #2 #3{{\sl Phys. Lett.} {\bf A #1} (#2) #3}
\def\PLB #1 #2 #3{{\sl Phys. Lett.} {\bf B #1} (#2) #3}
\def\JMP #1 #2 #3{{\sl J. Math. Phys.} {\bf #1} (#2) #3}
\def\PTP #1 #2 #3{{\sl Prog. Theor. Phys.} {\bf #1} (#2) #3}
\def\SPTP #1 #2 #3{{\sl Suppl. Prog. Theor. Phys.} {\bf #1} (#2) #3}
\def\AoP #1 #2 #3{{\sl Ann. of Phys.} {\bf #1} (#2) #3}
\def\PNAS #1 #2 #3{{\sl Proc. Natl. Acad. Sci. USA} {\bf #1} (#2) #3}
\def\RMP #1 #2 #3{{\sl Rev. Mod. Phys.} {\bf #1} (#2) #3}
\def\PR #1 #2 #3{{\sl Phys. Reports} {\bf #1} (#2) #3}
\def\AoM #1 #2 #3{{\sl Ann. of Math.} {\bf #1} (#2) #3}
\def\UMN #1 #2 #3{{\sl Usp. Mat. Nauk} {\bf #1} (#2) #3}
\def\FAP #1 #2 #3{{\sl Funkt. Anal. Prilozheniya} {\bf #1} (#2) #3}
\def\FAaIA #1 #2 #3{{\sl Functional Analysis and Its Application} {\bf
#1} (#2) #3}
\def\BAMS #1 #2 #3{{\sl Bull. Am. Math. Soc.} {\bf #1} (#2)
#3} \def\TAMS #1 #2 #3{{\sl Trans. Am. Math. Soc.} {\bf #1} (#2) #3}
\def\InvM #1 #2 #3{{\sl Invent. Math.} {\bf #1} (#2) #3}
\def\LMP #1 #2 #3{{\sl Letters in Math. Phys.} {\bf #1} (#2) #3}
\def\IJMPA #1 #2 #3{{\sl Int. J. Mod. Phys.} {\bf A #1} (#2) #3}
\def\AdM #1 #2 #3{{\sl Advances in Math.} {\bf #1} (#2) #3}
\def\RMaP #1 #2 #3{{\sl Reports on Math. Phys.} {\bf #1} (#2) #3}
\def\IJM #1 #2 #3{{\sl Ill. J. Math.} {\bf #1} (#2) #3}
\def\APP #1 #2 #3{{\sl Acta Phys. Polon.} {\bf #1} (#2) #3}
\def\TMP #1 #2 #3{{\sl Theor. Mat. Phys.} {\bf #1} (#2) #3}
\def\JPA #1 #2 #3{{\sl J. Physics} {\bf A#1} (#2) #3}
\def\JSM #1 #2 #3{{\sl J. Soviet Math.} {\bf #1} (#2) #3}
\def\MPLA #1 #2 #3{{\sl Mod. Phys. Lett.} {\bf A #1} (#2) #3}
\def\JETP #1 #2 #3{{\sl Sov. Phys. JETP} {\bf #1} (#2) #3}
\def\JETPL #1 #2 #3{{\sl  Sov. Phys. JETP Lett.} {\bf #1} (#2) #3}
\def\PHSA #1 #2 #3{{\sl Physica} {\bf A #1} (#2) #3}
\def\CQG #1 #2 #3{{\sl Class. Quantum Grav.} {\bf #1} (#2) #3}
\def\SJNP #1 #2 #3{{\sl Sov. J. Nucl. Phys. (Yadern.Fiz.)} {\bf #1} (#2) #3}
\def\a{\alpha}\def\b{\beta}\def\g{\gamma}

\def\k{\kappa}
\def\om{\omega}\def\Om{\Omega}

\newcommand{\nn}{\nonumber\\}\newcommand{\p}[1]{(\ref{#1})}
\begin{document}
\renewcommand{\thefootnote}{\fnsymbol{footnote}}
\thispagestyle{empty}
\begin{flushright}
Preprint DFPD 95/TH/58\\
hep-th/9510220\\
October 1995
\end{flushright}

\medskip
\begin{center}
{\large\bf
New supersymmetric generalization
of the  Liouville equation}\footnote{Work supported  in  part  by  the
International Science Foundation under the grant N RY 9200,
 by the State  Committee  for  Science  and  Technology  of
Ukraine under the Grant N 2/100 and
 by the INTAS grants 93--127, 93--493, 93--633, 94-2317}

\vspace{0.3cm}

{\bf Igor A. Bandos,}

\vspace{0.2cm}
{\it Kharkov Institute of Physics and Technology}
{\it 310108, Kharkov,  Ukraine}\\
e-mail:  kfti@rocket.kharkov.ua

\vspace{0.2cm}
\renewcommand{\thefootnote}{\dagger}

{\bf Dmitrij Sorokin}\footnote{on leave from Kharkov Institute of
Physics and Technology, Kharkov, 310108, Ukraine.},
\renewcommand{\thefootnote}{\ddagger}

\vspace{0.2cm}
{\it Universit\`a Degli Studi Di Padova
Dipartimento Di Fisica ``Galileo Galilei''\\
ed INFN, Sezione Di Padova
Via F. Marzolo, 8, 35131 Padova, Italia}\\
e--mail: sorokin@pd.infn.it

\vspace{0.2cm}
{\bf and}

\vspace{0.2cm}

{\bf Dmitrij V. Volkov}

\vspace{0.2cm}
{\it Kharkov Institute of Physics and Technology}
{\it 310108, Kharkov,  Ukraine}\\
e-mail:  kfti@rocket.kharkov.ua

\vspace{0.5cm}
{\bf Abstract}

\end{center}

We present new $n=(1,1)$ and $n=(1,0)$ supersymmetric generalization
of the Liouville equation, which originate from a geometrical
approach to describing the classical dynamics of  Green--Schwarz
superstrings in $N=2,~D=3$ and $N=1,~D=3$ target superspace. Considered
are a zero curvature representation and B\"acklund transformations
associated with the supersymmetric non--linear equations.

\bigskip


\setcounter{page}1
\renewcommand{\thefootnote}{\arabic{footnote}}
\setcounter{footnote}0

\newpage
\section{Introduction}
The dynamics of some physical systems is described by non--linear
equations, which are exactly solvable. Solitons, monopoles, instantons
and relativistic strings are examples of such systems.

For instance,
the classical dynamics of a free relativistic string in D=3 space--time
is governed by the Liouville equation \cite{lr,om,barnes}
and it is described by the
complex Liouville equation in D=4 space--time \cite{bnc,barnes}.
The Lagrangian, from which the
Liouville equation can be obtained  arises also
when one calculates the conformal anomaly in the quantum theory of
non--critical ($D \not = 26$) bosonic strings \cite{polya}. The calculation
of the superconformal anomaly of a relativistic spinning string in
$D\not = 10$ results in an effective Lagrangian, which produces
an exactly solvable $n=(1,1)$ \footnote{We denote the number
of left and right supersymmetries on the worldsheet by small $n$,
and by capital $N$ the number of supersymmetries in target space}
supersymmetric generalization of the
Liouville equation \cite{polya1} considered earlier in \cite{kulish}.
There has also been an activity in looking for and studying other
completely--integrable supersymmetric systems (see
\cite{f}--\cite{ft} and refs. therein).
As further development
of this research it seems of interest to find and analyse supersymmetric
non--linear equations that describe Green--Schwarz superstrings.

The purpose of this letter is to present an $n=(1,0)$ and $n=(1,1)$
worldsheet supersymmetric generalization of the Liouville equation which
describe the classical dynamics of Green--Schwarz superstrings in $N=1$,
$D=3$ and $N=2$, $D=3$ target superspace, respectively. The worldsheet
supersymmetry is a counterpart of the fermionic $\k$--symmetry of the
Green--Schwarz superstrings in a doubly--supersymmetric approach (see
\cite{bpstv} and refs. therein).

The $n=(1,1)$ super--Liouville equation we obtained turns out to be
different from and cannot be transformed  into the conventional one
\cite{kulish}
by any local change of variables. The superfield form of this new $n=(1,1)$
superfield generalization of the Liouville equation admits B\"acklund
transformations of variables, and
is exactly solvable in a ``trivial''
way, since at the component level the bosonic and fermionic part of the
$n=(1,1)$ superfield equation are reduced, respectively, to
the bosonic Liouville equation and the free
chiral fermion equations, which are completely disconnected from each
other. We would like to thank F. Toppan for pointing us this
reduction.

As to the $n=(1,0)$ case, it can be obtained as a reduction of both,
the standard and the new $n=(1,1)$ supersymmetric Liouville system.

To begin with, in Section 2 we shall briefly discuss how the Liouville
equation is extracted from the equations of motion of a bosonic string
in a form of the zero curvature representation, which allows one to
naturally get B\"acklund transformations for the Liouville variable. In
Section 3 we shall apply the same technique for analysing the new
$n=(1,1)$ and $n=(1,0)$ supersymmetric generalization
of the Liouville equation.

\section{The Liouville equation and the $D=3$ bosonic string}

The reduction of the string equations of motion to the Liouville
equation is achieved by changing string variables and solving for the
Virasoro constraints in such a way that only the variables which
correspond to the independent physical degrees of freedom of the string
\footnote{Remind that the bosonic string has one independent
transversal degree of freedom in $D=3$ and two in $D=4$}
remain and the latter satisfy the Liouville equation
\cite{lr,om,barnes,zhelt}:
\begin{equation}\label{leq}
\partial_{++} \partial_{--} w = exp\{ 2w \},
\end{equation}
where
$\partial_{\pm\pm} \equiv \partial / \partial \xi^{\pm\pm}$,
$\xi^{\pm\pm} = 2^{-1/2} (x^0 \pm x^1 )$ are light--cone coordinates on
the worldsheet, and the worldsheet function $w(\xi)$ is real for the
$D=3$ bosonic string and complex for the $D=4$ bosonic string.
(Each plus (minus) denotes the left (right)
Majorana--Weyl spinor representation of $SO(1,1)$, and the couples of
pluses and minuses denote the $SO(1,1)$ vector representation).

The equations of motion of a bosonic string in space--time of dimension
greater than four were transformed into a system of non--linear equations
in \cite{zhelt,barnes}.

A natural way of getting the nonlinear equations from the string
equations is to use a geometrical approach (\cite{lr,om,barnes,zhelt}
and references therein) which is based on the theory of surfaces
embedded into a target space. From the geometrical point of view the
functions which
enter the non--linear equations determine the geometrical objects
on the worldsheet, such as a metric and connection forms, induced by the
embedding. For instance, the $SO(1,1)$ spin connection has the form
\begin{equation}\label{Om0}
\Om^{(0)} = (d\xi^{++} \partial_{++}
- d\xi^{--} \partial_{--}) w - dl ,
\end{equation}
where $l(\xi)$ corresponds to a local $SO(1,1)$ transformation in the
space tangent to the worldsheet \footnote{ In \p{Om0}
and below the the external differential
$d$ and external product
are used in such a way that for a $p$--form   $\om_p$ and a $q$--form
$\om_q$ we have
$ \om_p \om_q = (-1)^{pq} \om_q \om_p,
d(\om_p \om_q) = \om_p d\om_q + (-1)^{q} d\om_p \om_q  $}.

In the geometrical approach  the form \p{Om0} together with
another two differential one--forms
\begin{equation}\label{Ompm}
\Om^{\pm\pm} =-2 d\xi^{\mp\mp} exp \{ w \mp l \},
\end{equation}
is part of an $SL(2,{\bf R})$ connection
\begin{equation}\label{Oms}
\Om_{\underline \a}^{~\underline \b} =  {1 \over 2}
\left(
      \matrix{ \Om^{(0)} & \Om^{--}  \cr
               \Om^{++} & - \Om^{(0)} \cr}
                                          \right)
\qquad ({\underline\a},{\underline\b}=1,2),
\end{equation}
which satisfies the $SL(2,{\bf R})$ Maurer--Cartan equation
\begin{equation}\label{MCs}
d\Om_{\underline \a}^{~\underline \b} -
\Om_{\underline \a}^{~\underline \g} \Om_{\underline \g}^{~\underline \b}
= 0.
\end{equation}
Note that $SL(2,{\bf R})$ is isomorphic to the Lorentz group in $D=3$
target space.

Eq. \p{MCs} contains the Liouville equation \p{leq} as the differential
form equation
\begin{equation}\label{G}
d\Om^{(0)} - 1/2 ~ \Om^{--} \Om^{++} = 0,
\end{equation}
called the Gauss equation in surface theory (see \cite{lr,om,barnes}).

Another two equations, which enter \p{MCs}, are so called
Peterson--Codazzi equations
\begin{equation}\label{PC}
d\Om^{\mp\mp} \mp \Om^{(0)} \Om^{\mp\mp} = 0,
\end{equation}
 which are identically satisfied  when $\Om^{(0)},~~\Om^{\mp\mp}$,
have the form \p{Om0}, \p{Ompm}.

 The system of equations \p{MCs} (or \p{G}, \p{PC}) describes
the classical motion of the $D=3$ bosonic string in the geometrical
approach \cite{lr,om,barnes}. It has the form of the zero curvature
condition for the connection \p{Oms} and can be considered as the
integrability condition for an associated  system of linear equations
\begin{equation}\label{lin}
d\Psi_{\underline \a}
- \Om_{\underline \a}^{~\underline \b}
\Psi_{\underline \b} = 0.
\end{equation}
The invariance of the theory under the local $SO(1,1)$  transformations,
which is reflected in the presence in eqs. \p{Om0}, \p{Ompm} of an
arbitrary field $l(\xi)$, allows one to reproduce a B\"acklund
transformation \cite{das,kulish,iv&kr} which relates a solution of the
Liouville equation \p{leq} with a solution of the free field equation
\begin{equation}\label{free}
\partial_{++} \partial_{--}{l} = 0.
\end{equation}
This is achieved by further specifying the $\Om^{(0)}$ form in \p{Om0}
by requiring its $d\xi^{++},~d\xi^{--}$ coefficients to be,
respectively,
\begin{eqnarray}\label{back}
\partial_{++} w - \partial_{++} l = {1 \over {b}} exp\{ w+l\} , \qquad
\nn
\partial_{--} w + \partial_{--} l = {b} exp\{ w-l\} , \qquad
\end{eqnarray}
where $b$ is an arbitrary constant.

Eqs. \p{back} are the B\"acklund transformations being a useful tool in
studying exactly solvable systems (\cite{kulish,das} and references
therein). The self--consistency conditions for \p{back} are the
Liouville equation \p{leq} for $w(\xi)$ and eq. \p{free} for $l(\xi)$.

Let us turn now to the  consideration of an $n=(1,0)$ and $n=(1,1)$
supersymmetric generalization of the Liouville equation.

\section{Worldsheet supersymmetric generalization of the Liouville
equation and $D=3$ Green--Schwarz superstrings}
Because of the Virasoro conditions and fermionic constraints related to
the local $\k$--symmetry of the Green--Schwarz strings \cite{gsw}
the classical dynamics of superstrings in $D=3$ superspace with N=1,2
Grassmann spinor coordinates is described by one independent bosonic
variable and one (or two) independent fermionic variables, respectively.
A doubly supersymmetric generalization \cite{bpstv} of
the geometrical approach \cite{lr,om,barnes} to the
consideration of superstrings and supermembranes as supersurfaces
embedded into a target superspace
allows one to reduce $N=1,2$, $D=3$ Green--Schwarz superstring equations
of motion to a system of equations for the independent superstring
variables \footnote{the details  of this reformulation will be published
elsewhere}. In the case of the $N=1$, $D=3$ Green--Schwarz superstring we
get the Liouville equation for the bosonic string variable $w(\xi)$ and
a chirality condition for the fermionic string variable $\psi_L(\xi):$
\begin{equation}\label{n1}
\partial_{++} \partial_{--} w = exp\{ 2w \},
\end{equation}
$$
\partial_{--}\psi_L=0.
$$

In the case of the $N=2$, $D=3$ Green--Schwarz superstring the system of
equations for the bosonic variable $w(\xi)$ and fermionic variables
$\psi_R(\xi)$, $\psi_L(\xi)$ of opposite worldsheet spinor chirality has the
form
\begin{equation}\label{n2}
\partial_{++} \partial_{--} w = exp\{ 2w \},
\end{equation}
$$
\partial_{++}\psi_R=0, \qquad \partial_{--}\psi_L=0.
$$
Eqs. \p{n2} reduce to \p{n1} when $\psi_R\equiv 0$.
The integrability of eqs. \p{n1}, \p{n2} is obvious.

Eqs. \p{n1} and \p{n2} possess, respectively, $n=(1,0)$ and $n=(1,1)$
supersymmetry (which replace $\k$--symmetry) on the worldsheet,
and, in fact, were derived from a
superfield system of equations which we shall write down below.

But firstly, let us present the conventional
$n=(1,1)$ supersymmetric Liouville
equation \cite{kulish}
\footnote{$n=(2,2)$ and $n=(4,4)$ supersymmetric generalization of the
Liouville equation were constructed in \cite{iv&kr}}
which arises in the quantum theory of
non--critical fermionic strings \cite{polya}:
\begin{equation}\label{kul}
D_{-} D_{+}\Phi = 2ie^\Phi,
\end{equation}
where
$
\Phi = \phi + i \eta^{+} {\psi}_++ i\eta^-\psi_-+ i\eta^{+}
\eta^{-} F
$ is a superfield in $n=(1,1)$ worldsheet superspace parametrized by
bosonic coordinates $\xi^{\pm\pm}$ and fermionic coordinates
$(\eta^+,\eta^-)$, and $D_{\pm} = {\partial \over {\partial \eta^{\pm}}}
+2i \eta^{\pm} \partial_{\pm\pm}$ are supercovariant derivatives which
form the flat $n=(1,1)$ superalgebra
\begin{equation}\label{alg}
D_{+} D_{+} = 2i \partial_{++}, \qquad
\{ D_{+}, D_{-} \} = 0 , \qquad
D_{-} D_{-} = 2i \partial_{--}.
\end{equation}
{}From eq. \p{kul} one gets the following system of equations for the
components of $\Phi(\xi,\eta)$
$$
\partial_{++}\partial_{--}\phi=-e^\phi(e^\phi+{1\over 2}i\psi_-\psi_+),
\qquad F=2e^\phi,
$$
\begin{equation}\label{kulik}
\partial_{++}\psi_-=- e^\phi\psi_+, \qquad
\partial_{--}\psi_+=e^\phi\psi_-.
\end{equation}

It is obvious that eqs. \p{n2} and \p{kulik} cannot be transformed
into each other by any
local change of variables. However, eqs. \p{n1} can be obtained from
\p{kulik} by a truncation of the latter to an $n=(1,0)$ supersymmetric
system. This truncation is performed by imposing an additional condition
on $e^{-\phi} \psi_-\equiv\psi_L$ to be a chiral field
$\partial_{--}\psi_L=0$. Then from \p{kulik} it follows that $\psi_+$ is
not an independent field anymore since its equation of motion becomes a
consequence of the two other.
identically satisfied. Thus, we get \p{n1} upon redefining $\phi\to
\phi^{\prime}=\phi-{i\over 4}\psi_L\partial_{++}\psi_L$ and replacing
$\partial_{--}\to-\partial_{--}$.

In contrast to \p{kul} the bosonic and fermionic variables of \p{n1} and
\p{n2} do not form a single superfield and transform non--lineary under
supersymmetry. Each of them (upon some field
redefinition) is the leading component of a corresponding bosonic
$W(\xi,\eta)$ or fermionic $\Psi_L(\xi,\eta)$, $\Psi_R(\xi,\eta)$
superfields.

$\Psi_L(\xi,\eta)$ and $\Psi_R(\xi,\eta)$ are chiral superfields:
\begin{equation}\label{left}
D_{-} \Psi_L = 0 ~~~ \qquad \Rightarrow  ~~~ \qquad
\partial_{--} \Psi_L = 0
\end{equation}
\begin{equation}\label{right}
D_{+} \Psi_R = 0 ~~~ \qquad \Rightarrow  ~~~ \qquad
\partial_{++} \Psi_R = 0,
\end{equation}
and are connected with $W(\xi,\eta)$ through the following relations:
\begin{equation}\label{constrl}
D_+ (e^{3W} \Psi_L) = e^{3W} ~~~ \qquad
\Rightarrow  ~~~ \qquad
D_+ \Psi_L + 3 D_+ W \Psi_L = 1 \qquad
\end{equation}
\begin{equation}\label{constrr}
D_- (e^{3W} \Psi_R) = e^{3W} ~~~ \qquad
\Rightarrow  ~~~ \qquad
D_- \Psi_R + 3 D_- W \Psi_R = 1.
\end{equation}
The  superfield form of eq. \p{n2} in $n=(1,1)$
worldsheet superspace $(\xi^{\pm\pm},\eta^+,\eta^-)$ is
\begin{equation}\label{sleq}
D_- D_+ W = 4e^{2W} \Psi_L \Psi_R;
\end{equation}
and the superfield form of \p{n1} in $n=(1,0)$  worldsheet superspace
$(\xi^{\pm\pm},\eta^+)$ is
\begin{equation}\label{1Nsleq}
\partial_{--} D_+ W = 2i e^{2W} \Psi_L
\end{equation}
where $W$ and $\Psi_L$ are taken to be independent of $\eta_-$
and $\Psi_R$ is put equal to zero.

The conditions \p{left}--\p{constrr} can be explicitly solved with
$W$, $\Psi_L$ and $\Psi_R$ having the following component form
\begin{equation}\label{wexp}
W  = w
+{{2i} \over 3} \eta^{+} e^{-3w} \partial_{++} (e^{3w}\psi_L)
+ {{2i} \over 3} \eta^{-} e^{-3w} \partial_{--} (e^{3w}\psi_R)
+ 4 \eta^+ \eta^- \psi_L \psi_R\partial_{++}\partial{--}w,
\end{equation}
\begin{equation}\label{psilexp}
\Psi_L  = \psi_L (\xi^{++}) +
\eta^{+} (1 -2i \partial_{++} \psi_L \psi_L) ,
\end{equation}
\begin{equation}\label{psirexp}
\Psi_R  = \psi_R (\xi^{--}) +
\eta^{-} (1 -2i \partial_{--} \psi_R \psi_R).
\end{equation}

In the doubly--supersymmetric geometrical approach the system of
equations \p{left}--\p{sleq}  arises as
the zero curvature condition \p{MCs} for
an $SL(2,{\bf
R})$ connection composed out of differential one--superforms
\begin{equation}\label{2NOm--}
\Om^{--} = \exp({W+L}) ( - 2 e^{++} (1 -  D_+ W \Psi_L) -
4i e^+ \Psi_L ) ,
\end{equation}
\begin{equation}\label{2NOm++}
\Om^{++} = \exp({W-L}) ( - 2 e^{--} (1 -  D_- W \Psi_R) -
4i e^- \Psi_R ) ,
\end{equation}
\begin{equation}\label{2NOm0}
\Om^{(0)} =
(e^{+} D_+ + e^{++} \partial_{++}
- e^{-} D_- - e^{--} \partial_{--}) W - dL,
\end{equation}
and the system of equations \p{left}, \p{constrl},
\p{1Nsleq} is part of
the zero curvature condition \p{MCs} for $SL(2,{\bf R})$
one--superforms
\begin{equation}\label{1NOm--}
\Om^{--} = \exp({W+L}) ( - 2 e^{++} (1 -  D_+ W \Psi_L) -
4i e^+ \Psi_L ) ,
\end{equation}
\begin{equation}\label{1NOm++}
\Om^{++} =  -2e^{--} \exp({W-L})
\end{equation}
\begin{equation}\label{1NOm0}
\Om^{(0)} =
(e^{+} D_+ + e^{++} \partial_{++}
-  e^{--} \partial_{--})W - dL.
\end{equation}
In \p{2NOm--}--\p{1NOm0}
$$
e^{\pm}\equiv d\eta^\pm,\qquad e^{++}\equiv
d\xi^{++}-2id\eta^+\eta^+,\qquad e^{--}\equiv
d\xi^{--}-2id\eta^-\eta^- ~~~({\rm for}~~ n=(1,1))
$$
$$
\hfill e^{--}\equiv d\xi^{--}~~~({\rm for}~~ n=(1,0))
$$

are the basic supercovariant one--forms in $d=2$ superspace.

One can check that the system of $n=(1,1)$ superfield equations
\p{left}--\p{sleq} is part of the Maurer--Cartan equations \p{MCs} for
the forms \p{2NOm--}--\p{2NOm0}, and
the system of $n=(1,0)$ superfield equations
\p{left}, \p{constrl}, \p{1Nsleq} is part of
the Maurer--Cartan equations \p{MCs} for
the forms \p{1NOm--}--\p{1NOm0}.

In the geometrical approach to Green--Schwarz superstrings \cite{bpstv}
the Maurer--Cartan equations for the forms \p{2NOm--}--\p{2NOm0} and
\p{1NOm--}--\p{1NOm0} describe, respectively, the classical motion of
the
$N=2$ and $N=1$, $D=3$ superstring, the $\k$--symmetry being replaced
by $n=(1,1)$ and $n=(1,0)$ worldsheet supersymmetry, respectively.

$n=(1,1)$  B\"acklund transformations
\begin{eqnarray}\label{2Nback}
D_{+} W - D_{+} L =
{{2i} \over {b}} exp\{ W+L\} \Psi_L , \qquad \nn
D_{-} W +
D_{-} L = {{2ib}}  exp\{ W-L\}\Psi_R,
\end{eqnarray}
and $n=(1,0)$  B\"acklund transformations
\begin{eqnarray}\label{1Nback}
D_{+} W - D_{+} L =
{{2i} \over {b}} exp\{ W+L\} \Psi_L , \qquad \nn
\partial_{--} W +
\partial_{--} L = b~exp\{ W-L\},
\end{eqnarray}
which relate the superfields $W$ and $L$ are obtained
as additional conditions imposed on the
components of $\Om^{(0)}$ forms \p{2NOm0}, \p{1NOm0}.

The integrability conditions for \p{2Nback} and \p{1Nback}
are, respectively, the non--linear equations \p{sleq}
and \p{1Nsleq} for $W$, and the linear superfield equations
for $L$:
\begin{equation}\label{L}
D_+D_-L=0~~({\rm for}~~n=(1,1)); \qquad D_+\partial_{--}L=0
{}~~({\rm for}~~n=(1,0)).
\end{equation}
For checking this one has to take into account eqs.
\p{left}--\p{constrr}.

\section{Conclusion and discussion}

We have presented the  new exactly solvable
$n=(1,1)$ and $n=(1,0)$ supersymmetric
generalization of the Liouville equation, which originate from
the geometrical approach to describing $N=2,~1$,  superstrings in $D=3$
\cite{bpstv}. At the component level they split into the purely bosonic
Liouville equation and the free fermion equations.

The $n=(1,0)$ supersymmetric system of eqs. \p{n1} (or \p{left},
\p{constrl}, \p{1Nsleq}) can be obtained (by a truncation) of both, the
standard $n=(1,1)$ supersymmetric generalization \p{kulik} (or \p{kul})
and the new $n=(1,1)$ supersymmetric generalization
\p{n2} (or \p{left}--\p{sleq}) of the Liouville equation.

However, the two $n=(1,1)$ supersymmetric versions of the Liouville
equation are not connected with each other by any local transformation
of variables. To see this one can check, that (without reducing the
model to n=(1,0) case) it is not possible, by use
of local operations, to
construct chiral fermions from the components (or the superfield $W$)
of the standard supersymmetric Liouville equation \p{kulik}, \p{kul},
while the chiral fermions are part of the new $n=(1,1)$ supersymmetric
system of eqs. \p{n2}, \p{left}--\p{sleq}. The indirect way to connect
the two equations is
to make the B\"acklund transformation
\p{2Nback} from a solution of the new super--Liouville equation
to a solution of the free differential equation \p{L} and then to
make the B\"acklund transformation from the solution of \p{L} to a
solution of the standard super--Liouville equation \p{kul} \footnote{we
thank E. Ivanov for pointing our attention to this relationship}.

The problem of the relationship between the two $n=(1,1)$ supersymmetric
non--linear systems seems to be also
connected with the problem of classical (and
quantum) equivalence of (non--critical) Green--Schwarz and
relativistic spinning strings \cite{gsw,vz,berk}.

To trace this relationship one should compare the effective Liouville
Lagrangian originated from the quantum theory of spinning strings
\cite{polya1}
with an effective Lagrangian which should arise as a result of the
computation of anomalies in non--critical Green--Schwarz superstrings,
or compare the $n=(1,1)$ supersymmetric generalization of the
Liouville equation describing the classical $N=2, ~D=3$ Green--Schwarz
superstring with an $n=(1,1)$ supersymmetric system of equations to which
one should reduce the equations of motions of an $n=(1,1)$, $D=3$
spinning string. As far as we know, neither the effective Lagrangian for the
non--critical Green--Schwarz superstrings, nor the $n=(1,1)$
supersymmetric non--linear equations describing the classical dynamics
of the $n=(1,1)$, $D=3$ spinning string have been obtained yet.

Another difference between the standard and the new $n=(1,1)$
supersymmetric Liouville system is that the zero curvature
representation for the former is associated with an $OSp(1\vert 2)$
connection \cite{kulish,iv&kr,top},
while in the case considered above it is associated with a
connection of $SL(2,{\bf R})$ which is the structure group of the flat
$D=3$ target superspace.
This simpler group structure resulted in a ``trivial" component form \p{n2}
of our super--Liouville system in comparison with the standard one
\p{kulik}.

Thus, the two ways of supersymmetrizing the Liouville equation
are essencially different at this point and remind the situation with
the non--linear Schr\"edinger equation for which supersymmetric
versions based on $OSp(1\vert 2)$ \cite{popovich} and $SL(2,{\bf R})$
\cite{ft} are known to be different.

To conclude, we would like to note that the geometrical approach which
has been used above for obtaining non--linear equations describing
strings is applicable to studying p--branes ($p>1$) as well
\cite{barnes}.
The equations of motion of p--branes can also be rewritten as a zero
curvature representation treated as a self--consistency condition
for an associated system of linear equations.

In a recent paper \cite{hoppe95} non--linear equations of motion of a
bosonic ($D-2$)--brane moving in D--dimensional space--time were reduced
to a form of the zero curvature representation by fixing a gauge with
respect to the worldsurface diffeomorphisms and the Lorentz
transformations. In \cite{hoppe95} the zero curvature representation
and an associated system of linear equations obtained this way were
used, in particular, for finding explicit solutions and constructing
non--local conservation charges describing a classical motion of
$(D-2)$--branes.

The geometrical approach \cite{lr,om,barnes,zhelt,bpstv},
upon which the present letter is heavily based, produces a
Lorentz--covariant counterpart of the construction of ref. \cite{hoppe95}
for any bosonic p--branes.

\bigskip
\noindent
{\bf Acknowledgements}
The authors would like to thank  P. Howe, E. Ivanov,
S. Krivonos, M. Mukhtarov, P. Pasti, A. Sorin, M. Tonin and F. Toppan
for interest to this work and fruitful discussion. Special thanks are to
F. Toppan who helped us to analyse the component structure of the
non--linear equations.
I.B. and D.V. are grateful to Paolo Pasti and Mario Tonin for hospitality at
the Padova Section of the INFN and Physics Department of Padova
University, where part of this work was carried out.


\begin{thebibliography}{99}
\bibitem{lr}
F. Lund and T. Regge, \PRD 14 1976 1524.

\bibitem{om}
R. Omnes, \NPB 149 1979 269.

\bibitem{barnes}
B. M. Barbashov and V. V. Nesterenko, {\sl Introduction to the
relativistic string theory} World Scientific, 1990.

\bibitem{bnc}
B. M. Barbashov, V. V. Nesterenko and A. M. Chervyakov, {\sl J. Phys. A}
{\bf 13} (1980) 301.

\bibitem{polya}
A. M. Polyakov \PLB 103 1981 207.

\bibitem{polya1}
A. M. Polyakov \PLB 103 1981 210.

\bibitem{kulish}
M. Chaichian and P. P. Kulish, \PLB 78 1978 413.

\bibitem{f}
P. Di Vecchia and S. Ferrara, \NPB 130 1977 93;\\
S. Ferrara, L. Girardello and S. Sciuto, \PLB 76 1978 303;\\
L. Girardello and S. Sciuto, \PLB 77 (1978) 267.
\bibitem{h}
J. Hruby, \NPB 131 1977 275.
\bibitem{iv&kr}
E. A. Ivanov and S. O. Krivonos, {\sl Lett. Math. Phys.} {\bf 7} (1983)
523; {\sl Ibid} {\bf 8} (1984)
39, 345 (E); {\sl J. Phys.} {\bf A17} (1984) L671;\\
E. A. Ivanov, S. O. Krivonos and V. Leviant \NPB 304 1988 601.
\bibitem{top}
F. Toppan, \PLB 260 1991 346; \\
F. Toppan and Y.--Z. Zhang, \PLB 292 1992 67.

\bibitem{popovich}
P. P. Kulish, {\sl Lett. Math. Phys.} {\bf 10} (1985) 87.\\
Z. Popovich, {\sl Phys. Lett.} {\bf 194A} (1994) {375}.

\bibitem{ft}
F. Toppan, {\sl Int. J. Mod. Phys.} {\bf A10} (1995) 895.
\bibitem{bpstv}
I. Bandos, P. Pasti, D. Sorokin, M. Tonin and D. Volkov,
{\sl Nucl.Phys.} {\bf B446} (1995) 79, {\bf hep-th/9501113}.

\bibitem{zhelt}
A. Zheltukhin, \SJNP 33 1981 1723 ; \TMP 52 1982 73 ;
\PLB 116 1982 147; \TMP 56 1983 230.\\
A. N. Leznov and M. V. Saveliev, {\sl Commun. Math. Phys.} {\bf 89}
(1983) 59.\\
B. M. Barbashov, V. V. Nesterenko and A. M. Chervyakov,
\TMP 59 1984 209.

\bibitem{das}
A. Das, {\sl Integrable Models}, World Scientific, 1989.

\bibitem{int}
L. Faddeev and L. Takhtajan, {\sl ``Hamiltonian Methods in the Theory
of
Solitons''}, Springer (1987); \\
L. Dickey,{\sl ``Soliton Equations and Hamiltonian Systems''}, World
Scientific (1991); \\
N. Bogolyubov, A. Izergin and V. Korepin, {\sl ``Quantum Inverse
Scattering Method and Correlation Functions''}, Cambridge Univ. Press
(1993).

\bibitem{gsw}
M. Green, J. Schwarz and E. Witten, Superstring Theory, CUP, 1987
(and references therein).

\bibitem{vz}
D. V. Volkov and A. Zheltukhin, \JETPL 48 1988 61;
\LMP 17 1989 141; \NPB 335 1990 723.\\
D. Sorokin, V. Tkach, D. V. Volkov and A. Zheltukhin, {\sl Phys. Lett.}
{\bf B216} (1989) 302.\\
S. Aoyama, P. Pasti and M. Tonin, {\sl Phys. Lett.} {\bf
B283} (1992) 213.
\bibitem{berk}
N. Berkovits, \NPB 431 1994 258.
\bibitem{hoppe95}
J. Hoppe,
{\sl ``Conservation lows and formation of singularities in
relativistic
theories of extended objects''}, Preprint {\bf ETH--TH/95--7},
Zurich, February 1995, {\bf hep-th/9503069}.

\end{thebibliography}
\end{document}